%
%
%
%
%
%
%
%
\def\standardrisposta{s }\def\reducedrisposta{r }
\def\mplarisposta{mpla }\def\zerorisposta{z }\def\bigrisposta{big }
\def\doublerisposta{d }\def\cartarisposta{e }\def\amsrisposta{y }
\newcount\ingrandimento \newcount\sinnota \newcount\dimnota
\newcount\unoduecol \newdimen\collhsize \newdimen\tothsize
\newdimen\fullhsize \newcount\controllorisposta \sinnota=1
\newskip\infralinea  \global\controllorisposta=0
\immediate\write16 { ********  Welcome to PANDA macros (Plain TeX,
AP, 1991) ******** }
%
%
%
%
\def\risposta{s } 
\def\srisposta{u } 
\def\arisposta{y }
\ifx\risposta\standardrisposta \ingrandimento=1200
\message {>> This will come out UNREDUCED << }
\dimnota=2 \unoduecol=1 \global\controllorisposta=1 \fi
\ifx\risposta\bigrisposta \ingrandimento=1440
\message {>> This will come out ENLARGED << }
\dimnota=2 \unoduecol=1 \global\controllorisposta=1 \fi
\ifx\risposta\reducedrisposta \ingrandimento=1095 \dimnota=1
\unoduecol=1  \global\controllorisposta=1
\message {>> This will come out REDUCED << } \fi
\ifx\risposta\doublerisposta \ingrandimento=1000 \dimnota=2
\unoduecol=2  \message {>> You must print this in
LANDSCAPE orientation << } \global\controllorisposta=1 \fi
\ifx\risposta\mplarisposta \ingrandimento=1000 \dimnota=1
\message {>> Mod. Phys. Lett. A format << }
\unoduecol=1 \global\controllorisposta=1 \fi
\ifx\risposta\zerorisposta \ingrandimento=1000 \dimnota=2
\message {>> Zero Magnification format << }
\unoduecol=1 \global\controllorisposta=1 \fi
\ifnum\controllorisposta=0  \ingrandimento=1200
\message {>>> ERROR IN INPUT, I ASSUME STANDARD
UNREDUCED FORMAT <<< }  \dimnota=2 \unoduecol=1 \fi
\magnification=\ingrandimento
%
%
%
%
\newdimen\eucolumnsize \newdimen\eudoublehsize \newdimen\eudoublevsize
\newdimen\uscolumnsize \newdimen\usdoublehsize \newdimen\usdoublevsize
\newdimen\eusinglehsize \newdimen\eusinglevsize \newdimen\ussinglehsize
\newskip\standardbaselineskip \newdimen\ussinglevsize
\newskip\reducedbaselineskip \newskip\doublebaselineskip
\newskip\bigbaselineskip
\eucolumnsize=12.0truecm    
\eudoublehsize=25.5truecm   
\eudoublevsize=6.5truein    
\uscolumnsize=4.4truein     
\usdoublehsize=9.4truein    
\usdoublevsize=6.8truein    
\eusinglehsize=6.3truein    
\eusinglevsize=24truecm     
\ussinglehsize=6.5truein    
\ussinglevsize=8.9truein    
\bigbaselineskip=18pt plus.2pt       
\standardbaselineskip=16pt plus.2pt  
\reducedbaselineskip=14pt plus.2pt   
\doublebaselineskip=12pt plus.2pt    
%
%
\def\Portoffset{}
\def\Landoffset{\hoffset=-.140truein}
\ifx\risposta\mplarisposta \def\Portoffset{\hoffset=1.9truecm
\voffset=1.4truecm} \fi
%
%
\def\Landspec{}
\tolerance=10000
\parskip=0pt plus2pt  \leftskip=0pt \rightskip=0pt
%
%
\ifx\risposta\bigrisposta      \infralinea=\bigbaselineskip \fi
\ifx\risposta\standardrisposta \infralinea=\standardbaselineskip \fi
\ifx\risposta\reducedrisposta  \infralinea=\reducedbaselineskip \fi
\ifx\risposta\doublerisposta   \infralinea=\doublebaselineskip \fi
\ifx\risposta\mplarisposta     \infralinea=13pt \fi
\ifx\risposta\zerorisposta     \infralinea=12pt plus.2pt\fi
\ifnum\controllorisposta=0    \infralinea=\standardbaselineskip \fi
\ifx\risposta\doublerisposta   \Landoffset \else \Portoffset \fi
\ifx\risposta\doublerisposta \ifx\srisposta\cartarisposta
\tothsize=\eudoublehsize \collhsize=\eucolumnsize
\vsize=\eudoublevsize  \else  \tothsize=\usdoublehsize
\collhsize=\uscolumnsize \vsize=\usdoublevsize \fi \else
\ifx\srisposta\cartarisposta \tothsize=\eusinglehsize
\vsize=\eusinglevsize \else  \tothsize=\ussinglehsize
\vsize=\ussinglevsize \fi \collhsize=4.4truein \fi
\ifx\risposta\mplarisposta \tothsize=5.0truein
\vsize=7.8truein \collhsize=4.4truein \fi
%
%
%
%
\newcount\contaeuler \newcount\contacyrill \newcount\contaams \newcount\contasym
\font\ninerm=cmr9  \font\eightrm=cmr8  \font\sixrm=cmr6
\font\ninei=cmmi9  \font\eighti=cmmi8  \font\sixi=cmmi6
\font\ninesy=cmsy9  \font\eightsy=cmsy8  \font\sixsy=cmsy6
\font\ninebf=cmbx9  \font\eightbf=cmbx8  \font\sixbf=cmbx6
\font\ninett=cmtt9  \font\eighttt=cmtt8  \font\nineit=cmti9
\font\eightit=cmti8 \font\ninesl=cmsl9  \font\eightsl=cmsl8
\skewchar\ninei='177 \skewchar\eighti='177 \skewchar\sixi='177
\skewchar\ninesy='60 \skewchar\eightsy='60 \skewchar\sixsy='60
\hyphenchar\ninett=-1 \hyphenchar\eighttt=-1 \hyphenchar\tentt=-1
%
\font\tencmmib=cmmib10  \newfam\cmmibfam  \skewchar\tencmmib='177
\font\tencmbsy=cmbsy10  \newfam\cmbsyfam  \skewchar\tencmbsy='60
\def\scaps{\cmcsc}                 
\font\tencmcsc=cmcsc10  \newfam\cmcscfam
\ifnum\ingrandimento=1095

\font\capsone=cmcsc10 at 10.95pt 

\else

\font\capsone=cmcsc10 at 12pt 
\fi

\def\ttaarr{\bf}                
\def\ppaarr{\sl}                

%
%
%
\newfam\eufmfam \newfam\msamfam \newfam\msbmfam \newfam\eufbfam
\def\Loadeulerfonts{\global\contaeuler=1 \ifx\arisposta\amsrisposta
\font\teneufm=eufm10              
\font\eighteufm=eufm8 \font\nineeufm=eufm9 \font\sixeufm=eufm6
\font\seveneufm=eufm7  \font\fiveeufm=eufm5
\font\teneufb=eufb10              
\font\eighteufb=eufb8 \font\nineeufb=eufb9 \font\sixeufb=eufb6
\font\seveneufb=eufb7  \font\fiveeufb=eufb5
\font\teneurm=eurm10              
\font\eighteurm=eurm8 \font\nineeurm=eurm9
\font\teneurb=eurb10              
\font\eighteurb=eurb8 \font\nineeurb=eurb9
\font\teneusm=eusm10              
\font\eighteusm=eusm8 \font\nineeusm=eusm9
\font\teneusb=eusb10              
\font\eighteusb=eusb8 \font\nineeusb=eusb9
\else \def\eufm{\tt} \def\eufb{\tt} \def\eurm{\tt} \def\eurb{\tt}
\def\eusm{\tt} \def\eusb{\tt}    \fi}
\def\loadamsmath{\global\contaams=1 \ifx\arisposta\amsrisposta
\font\tenmsam=msam10 \font\ninemsam=msam9 \font\eightmsam=msam8
\font\sevenmsam=msam7 \font\sixmsam=msam6 \font\fivemsam=msam5
\font\tenmsbm=msbm10 \font\ninemsbm=msbm9 \font\eightmsbm=msbm8
\font\sevenmsbm=msbm7 \font\sixmsbm=msbm6 \font\fivemsbm=msbm5
\else \def\msbm{\bf} \fi \def\Bbb{\msbm} \def\symbl{\msam} \tenpoint}
\def\loadcyrill{\global\contacyrill=1 \ifx\arisposta\amsrisposta
\font\tenwncyr=wncyr10 \font\ninewncyr=wncyr9 \font\eightwncyr=wncyr8
\font\tenwncyb=wncyr10 \font\ninewncyb=wncyr9 \font\eightwncyb=wncyr8
\font\tenwncyi=wncyr10 \font\ninewncyi=wncyr9 \font\eightwncyi=wncyr8
\else \def\cyrill{\sl} \def\cyrilb{\sl} \def\cyrili{\sl} \fi\tenpoint}
\catcode`\@=11
\def\undefine#1{\let#1\undefined}
\def\newsymbol#1#2#3#4#5{\let\next@\relax
 \ifnum#2=\@ne\let\next@\msafam@\else
 \ifnum#2=\tw@\let\next@\msbfam@\fi\fi
 \mathchardef#1="#3\next@#4#5}
\def\mathhexbox@#1#2#3{\relax
 \ifmmode\mathpalette{}{\m@th\mathchar"#1#2#3}%
 \else\leavevmode\hbox{$\m@th\mathchar"#1#2#3$}\fi}
\def\hexnumber@#1{\ifcase#1 0\or 1\or 2\or 3\or 4\or 5\or 6\or 7\or 8\or 
9\or A\or B\or C\or D\or E\or F\fi}
\edef\msafam@{\hexnumber@\msamfam}
\edef\msbfam@{\hexnumber@\msbmfam}
\mathchardef\dabar@"0\msafam@39
\catcode`\@=12    
\def\loadamssym{\ifx\arisposta\amsrisposta  \ifnum\contaams=1 
\global\contasym=1 
\catcode`\@=11
\def\dashrightarrow{\mathrel{\dabar@\dabar@\mathchar"0\msafam@4B}}
\def\dashleftarrow{\mathrel{\mathchar"0\msafam@4C\dabar@\dabar@}}
\let\dasharrow\dashrightarrow
\def\ulcorner{\delimiter"4\msafam@70\msafam@70 }
\def\urcorner{\delimiter"5\msafam@71\msafam@71 }
\def\llcorner{\delimiter"4\msafam@78\msafam@78 }
\def\lrcorner{\delimiter"5\msafam@79\msafam@79 }
\def\yen{{\mathhexbox@\msafam@55}}
\def\checkmark{{\mathhexbox@\msafam@58 }}
\def\circledR{{\mathhexbox@\msafam@72 }}
\def\maltese{{\mathhexbox@\msafam@7A }}
\catcode`\@=12 
\input amssym.tex     \else  
\message{Panda error - First you have to use loadamsmath !!!!} \fi
\else \message{Panda error - You need the AMSFonts for these symbols 
!!!!}\fi}
\ifx\arisposta\amsrisposta
\font\sevenex=cmex7               
\font\eightex=cmex8  \font\nineex=cmex9
\font\ninecmmib=cmmib9   \font\eightcmmib=cmmib8
\font\sevencmmib=cmmib7 \font\sixcmmib=cmmib6
\font\fivecmmib=cmmib5   \skewchar\ninecmmib='177
\skewchar\eightcmmib='177  \skewchar\sevencmmib='177
\skewchar\sixcmmib='177   \skewchar\fivecmmib='177
\font\ninecmbsy=cmbsy9    \font\eightcmbsy=cmbsy8
\font\sevencmbsy=cmbsy7  \font\sixcmbsy=cmbsy6
\font\fivecmbsy=cmbsy5   \skewchar\ninecmbsy='60
\skewchar\eightcmbsy='60  \skewchar\sevencmbsy='60
\skewchar\sixcmbsy='60    \skewchar\fivecmbsy='60
\font\ninecmcsc=cmcsc9    \font\eightcmcsc=cmcsc8     \else
\def\cmmib{\fam\cmmibfam\tencmmib}\textfont\cmmibfam=\tencmmib
\scriptfont\cmmibfam=\tencmmib \scriptscriptfont\cmmibfam=\tencmmib
\def\cmbsy{\fam\cmbsyfam\tencmbsy} \textfont\cmbsyfam=\tencmbsy
\scriptfont\cmbsyfam=\tencmbsy \scriptscriptfont\cmbsyfam=\tencmbsy
\scriptfont\cmcscfam=\tencmcsc \scriptscriptfont\cmcscfam=\tencmcsc
\def\cmcsc{\fam\cmcscfam\tencmcsc} \textfont\cmcscfam=\tencmcsc \fi
\catcode`@=11
\newskip\ttglue
\gdef\tenpoint{\def\rm{\fam0\tenrm}
  \textfont0=\tenrm \scriptfont0=\sevenrm \scriptscriptfont0=\fiverm
  \textfont1=\teni \scriptfont1=\seveni \scriptscriptfont1=\fivei
  \textfont2=\tensy \scriptfont2=\sevensy \scriptscriptfont2=\fivesy
  \textfont3=\tenex \scriptfont3=\tenex \scriptscriptfont3=\tenex
  \def\mcal{\fam2 \tensy}  \def\mmit{\fam1 \teni}
  \textfont\itfam=\tenit \def\it{\fam\itfam\tenit}
  \textfont\slfam=\tensl \def\sl{\fam\slfam\tensl}
  \textfont\ttfam=\tentt \scriptfont\ttfam=\eighttt
  \scriptscriptfont\ttfam=\eighttt  \def\tt{\fam\ttfam\tentt}
  \textfont\bffam=\tenbf \scriptfont\bffam=\sevenbf
  \scriptscriptfont\bffam=\fivebf \def\bf{\fam\bffam\tenbf}
     \ifx\arisposta\amsrisposta    \ifnum\contaeuler=1
  \textfont\eufmfam=\teneufm \scriptfont\eufmfam=\seveneufm
  \scriptscriptfont\eufmfam=\fiveeufm \def\eufm{\fam\eufmfam\teneufm}
  \textfont\eufbfam=\teneufb \scriptfont\eufbfam=\seveneufb
  \scriptscriptfont\eufbfam=\fiveeufb \def\eufb{\fam\eufbfam\teneufb}
  \def\eurm{\teneurm} \def\eurb{\teneurb} \def\eusm{\teneusm}
  \def\eusb{\teneusb}    \fi    \ifnum\contaams=1
  \textfont\msamfam=\tenmsam \scriptfont\msamfam=\sevenmsam
  \scriptscriptfont\msamfam=\fivemsam \def\msam{\fam\msamfam\tenmsam}
  \textfont\msbmfam=\tenmsbm \scriptfont\msbmfam=\sevenmsbm
  \scriptscriptfont\msbmfam=\fivemsbm \def\msbm{\fam\msbmfam\tenmsbm}
     \fi      \ifnum\contacyrill=1     \def\cyrill{\tenwncyr}
  \def\cyrilb{\tenwncyb}  \def\cyrili{\tenwncyi}         \fi
  \textfont3=\tenex \scriptfont3=\sevenex \scriptscriptfont3=\sevenex
  \def\cmmib{\fam\cmmibfam\tencmmib} \scriptfont\cmmibfam=\sevencmmib
  \textfont\cmmibfam=\tencmmib  \scriptscriptfont\cmmibfam=\fivecmmib
  \def\cmbsy{\fam\cmbsyfam\tencmbsy} \scriptfont\cmbsyfam=\sevencmbsy
  \textfont\cmbsyfam=\tencmbsy  \scriptscriptfont\cmbsyfam=\fivecmbsy
  \def\cmcsc{\fam\cmcscfam\tencmcsc} \scriptfont\cmcscfam=\eightcmcsc
  \textfont\cmcscfam=\tencmcsc \scriptscriptfont\cmcscfam=\eightcmcsc
     \fi            \tt \ttglue=.5em plus.25em minus.15em
  \normalbaselineskip=12pt
  \setbox\strutbox=\hbox{\vrule height8.5pt depth3.5pt width0pt}
  \let\sc=\eightrm \let\big=\tenbig   \normalbaselines
  \baselineskip=\infralinea  \rm}
\gdef\ninepoint{\def\rm{\fam0\ninerm}
  \textfont0=\ninerm \scriptfont0=\sixrm \scriptscriptfont0=\fiverm
  \textfont1=\ninei \scriptfont1=\sixi \scriptscriptfont1=\fivei
  \textfont2=\ninesy \scriptfont2=\sixsy \scriptscriptfont2=\fivesy
  \textfont3=\tenex \scriptfont3=\tenex \scriptscriptfont3=\tenex
  \def\mcal{\fam2 \ninesy}  \def\mmit{\fam1 \ninei}
  \textfont\itfam=\nineit \def\it{\fam\itfam\nineit}
  \textfont\slfam=\ninesl \def\sl{\fam\slfam\ninesl}
  \textfont\ttfam=\ninett \scriptfont\ttfam=\eighttt
  \scriptscriptfont\ttfam=\eighttt \def\tt{\fam\ttfam\ninett}
  \textfont\bffam=\ninebf \scriptfont\bffam=\sixbf
  \scriptscriptfont\bffam=\fivebf \def\bf{\fam\bffam\ninebf}
     \ifx\arisposta\amsrisposta  \ifnum\contaeuler=1
  \textfont\eufmfam=\nineeufm \scriptfont\eufmfam=\sixeufm
  \scriptscriptfont\eufmfam=\fiveeufm \def\eufm{\fam\eufmfam\nineeufm}
  \textfont\eufbfam=\nineeufb \scriptfont\eufbfam=\sixeufb
  \scriptscriptfont\eufbfam=\fiveeufb \def\eufb{\fam\eufbfam\nineeufb}
  \def\eurm{\nineeurm} \def\eurb{\nineeurb} \def\eusm{\nineeusm}
  \def\eusb{\nineeusb}     \fi   \ifnum\contaams=1
  \textfont\msamfam=\ninemsam \scriptfont\msamfam=\sixmsam
  \scriptscriptfont\msamfam=\fivemsam \def\msam{\fam\msamfam\ninemsam}
  \textfont\msbmfam=\ninemsbm \scriptfont\msbmfam=\sixmsbm
  \scriptscriptfont\msbmfam=\fivemsbm \def\msbm{\fam\msbmfam\ninemsbm}
     \fi       \ifnum\contacyrill=1     \def\cyrill{\ninewncyr}
  \def\cyrilb{\ninewncyb}  \def\cyrili{\ninewncyi}         \fi
  \textfont3=\nineex \scriptfont3=\sevenex \scriptscriptfont3=\sevenex
  \def\cmmib{\fam\cmmibfam\ninecmmib}  \textfont\cmmibfam=\ninecmmib
  \scriptfont\cmmibfam=\sixcmmib \scriptscriptfont\cmmibfam=\fivecmmib
  \def\cmbsy{\fam\cmbsyfam\ninecmbsy}  \textfont\cmbsyfam=\ninecmbsy
  \scriptfont\cmbsyfam=\sixcmbsy \scriptscriptfont\cmbsyfam=\fivecmbsy
  \def\cmcsc{\fam\cmcscfam\ninecmcsc} \scriptfont\cmcscfam=\eightcmcsc
  \textfont\cmcscfam=\ninecmcsc \scriptscriptfont\cmcscfam=\eightcmcsc
     \fi            \tt \ttglue=.5em plus.25em minus.15em
  \normalbaselineskip=11pt
  \setbox\strutbox=\hbox{\vrule height8pt depth3pt width0pt}
  \let\sc=\sevenrm \let\big=\ninebig \normalbaselines\rm}
\gdef\eightpoint{\def\rm{\fam0\eightrm}
  \textfont0=\eightrm \scriptfont0=\sixrm \scriptscriptfont0=\fiverm
  \textfont1=\eighti \scriptfont1=\sixi \scriptscriptfont1=\fivei
  \textfont2=\eightsy \scriptfont2=\sixsy \scriptscriptfont2=\fivesy
  \textfont3=\tenex \scriptfont3=\tenex \scriptscriptfont3=\tenex
  \def\mcal{\fam2 \eightsy}  \def\mmit{\fam1 \eighti}
  \textfont\itfam=\eightit \def\it{\fam\itfam\eightit}
  \textfont\slfam=\eightsl \def\sl{\fam\slfam\eightsl}
  \textfont\ttfam=\eighttt \scriptfont\ttfam=\eighttt
  \scriptscriptfont\ttfam=\eighttt \def\tt{\fam\ttfam\eighttt}
  \textfont\bffam=\eightbf \scriptfont\bffam=\sixbf
  \scriptscriptfont\bffam=\fivebf \def\bf{\fam\bffam\eightbf}
     \ifx\arisposta\amsrisposta   \ifnum\contaeuler=1
  \textfont\eufmfam=\eighteufm \scriptfont\eufmfam=\sixeufm
  \scriptscriptfont\eufmfam=\fiveeufm \def\eufm{\fam\eufmfam\eighteufm}
  \textfont\eufbfam=\eighteufb \scriptfont\eufbfam=\sixeufb
  \scriptscriptfont\eufbfam=\fiveeufb \def\eufb{\fam\eufbfam\eighteufb}
  \def\eurm{\eighteurm} \def\eurb{\eighteurb} \def\eusm{\eighteusm}
  \def\eusb{\eighteusb}       \fi    \ifnum\contaams=1
  \textfont\msamfam=\eightmsam \scriptfont\msamfam=\sixmsam
  \scriptscriptfont\msamfam=\fivemsam \def\msam{\fam\msamfam\eightmsam}
  \textfont\msbmfam=\eightmsbm \scriptfont\msbmfam=\sixmsbm
  \scriptscriptfont\msbmfam=\fivemsbm \def\msbm{\fam\msbmfam\eightmsbm}
     \fi       \ifnum\contacyrill=1     \def\cyrill{\eightwncyr}
  \def\cyrilb{\eightwncyb}  \def\cyrili{\eightwncyi}         \fi
  \textfont3=\eightex \scriptfont3=\sevenex \scriptscriptfont3=\sevenex
  \def\cmmib{\fam\cmmibfam\eightcmmib}  \textfont\cmmibfam=\eightcmmib
  \scriptfont\cmmibfam=\sixcmmib \scriptscriptfont\cmmibfam=\fivecmmib
  \def\cmbsy{\fam\cmbsyfam\eightcmbsy}  \textfont\cmbsyfam=\eightcmbsy
  \scriptfont\cmbsyfam=\sixcmbsy \scriptscriptfont\cmbsyfam=\fivecmbsy
  \def\cmcsc{\fam\cmcscfam\eightcmcsc} \scriptfont\cmcscfam=\eightcmcsc
  \textfont\cmcscfam=\eightcmcsc \scriptscriptfont\cmcscfam=\eightcmcsc
     \fi             \tt \ttglue=.5em plus.25em minus.15em
  \normalbaselineskip=9pt
  \setbox\strutbox=\hbox{\vrule height7pt depth2pt width0pt}
  \let\sc=\sixrm \let\big=\eightbig \normalbaselines\rm }
\gdef\tenbig#1{{\hbox{$\left#1\vbox to8.5pt{}\right.\n@space$}}}
\gdef\ninebig#1{{\hbox{$\textfont0=\tenrm\textfont2=\tensy
   \left#1\vbox to7.25pt{}\right.\n@space$}}}
\gdef\eightbig#1{{\hbox{$\textfont0=\ninerm\textfont2=\ninesy
   \left#1\vbox to6.5pt{}\right.\n@space$}}}
\def\alternativefont#1#2{\ifx\arisposta\amsrisposta \relax \else
\xdef#1{#2} \fi}
\global\contaeuler=0 \global\contacyrill=0 \global\contaams=0
%
%
%
%
\newbox\fotlinebb \newbox\hedlinebb \newbox\leftcolumn
\gdef\makeheadline{\vbox to 0pt{\vskip-22.5pt
     \fullline{\vbox to8.5pt{}\the\headline}\vss}\nointerlineskip}
\gdef\makehedlinebb{\vbox to 0pt{\vskip-22.5pt
     \fullline{\vbox to8.5pt{}\copy\hedlinebb\hfil
     \line{\hfill\the\headline\hfill}}\vss} \nointerlineskip}
\gdef\makefootline{\baselineskip=24pt \fullline{\the\footline}}
\gdef\makefotlinebb{\baselineskip=24pt
    \fullline{\copy\fotlinebb\hfil\line{\hfill\the\footline\hfill}}}
\gdef\doubleformat{\shipout\vbox{\Landspec\makehedlinebb
     \fullline{\box\leftcolumn\hfil\columnbox}\makefotlinebb}
     \advancepageno}
\gdef\columnbox{\leftline{\pagebody}}
\gdef\line#1{\hbox to\hsize{\hskip\leftskip#1\hskip\rightskip}}
\gdef\fullline#1{\hbox to\fullhsize{\hskip\leftskip{#1}%
\hskip\rightskip}}
\gdef\footnote#1{\let\@sf=\empty
         \ifhmode\edef\#sf{\spacefactor=\the\spacefactor}\/\fi
         #1\@sf\vfootnote{#1}}
\gdef\vfootnote#1{\insert\footins\bgroup
         \ifnum\dimnota=1  \eightpoint\fi
         \ifnum\dimnota=2  \ninepoint\fi
         \ifnum\dimnota=0  \tenpoint\fi
         \interlinepenalty=\interfootnotelinepenalty
         \splittopskip=\ht\strutbox
         \splitmaxdepth=\dp\strutbox \floatingpenalty=20000
         \leftskip=\oldssposta \rightskip=\olddsposta
         \spaceskip=0pt \xspaceskip=0pt
         \ifnum\sinnota=0   \textindent{#1}\fi
         \ifnum\sinnota=1   \item{#1}\fi
         \footstrut\futurelet\next\fo@t}
\gdef\fo@t{\ifcat\bgroup\noexpand\next \let\next\f@@t
             \else\let\next\f@t\fi \next}
\gdef\f@@t{\bgroup\aftergroup\@foot\let\next}
\gdef\f@t#1{#1\@foot} \gdef\@foot{\strut\egroup}
\gdef\footstrut{\vbox to\splittopskip{}}
\skip\footins=\bigskipamount
\count\footins=1000  \dimen\footins=8in
\catcode`@=12
\tenpoint
\ifnum\unoduecol=1 \hsize=\tothsize   \fullhsize=\tothsize \fi
\ifnum\unoduecol=2 \hsize=\collhsize  \fullhsize=\tothsize \fi
\global\let\lrcol=L      \ifnum\unoduecol=1
\output{\plainoutput{\ifnum\tipbnota=2 \clearnmbnota\fi}} \fi
\ifnum\unoduecol=2 \output{\if L\lrcol
     \global\setbox\leftcolumn=\columnbox
     \global\setbox\fotlinebb=\line{\hfill\the\footline\hfill}
     \global\setbox\hedlinebb=\line{\hfill\the\headline\hfill}
     \advancepageno  \global\let\lrcol=R
     \else  \doubleformat \global\let\lrcol=L \fi
     \ifnum\outputpenalty>-20000 \else\dosupereject\fi
     \ifnum\tipbnota=2\clearnmbnota\fi }\fi
\def\ifdoublepage{\ifnum\unoduecol=2 }
\gdef\yespagenumbers{\footline={\hss\tenrm\folio\hss}}
\gdef\ciao{ \ifnum\fdefcontre=1 \endfdef\fi
     \par\vfill\supereject \ifnum\unoduecol=2
     \if R\lrcol  \headline={}\nopagenumbers\null\vfill\eject
     \fi\fi \end}

\newskip\olddsposta \newskip\oldssposta
\global\oldssposta=\leftskip \global\olddsposta=\rightskip

\def\filldots{\leaders\hbox to 1em{\hss.\hss}\hfill}
\def\inquadrb#1 {\vbox {\hrule  \hbox{\vrule \vbox {\vskip .2cm
    \hbox {\ #1\ } \vskip .2cm } \vrule  }  \hrule} }
 \def\newline{\hfil\break}
\def\jump{\vskip\baselineskip} \newskip\iinnffrr
\def\sjump{\iinnffrr=\baselineskip
          \divide\iinnffrr by 2 \vskip\iinnffrr}
\def\bjump{\vskip\baselineskip \vskip\baselineskip}
\newcount\nmbnota  \def\clearnmbnota{\global\nmbnota=0}
\newcount\tipbnota \def\letterfootnote{\global\tipbnota=1}

\def\note#1{\global\advance\nmbnota by 1 \ifnum\tipbnota=1
    \footnote{$^{\rm\nttlett}$}{#1} \else {\ifnum\tipbnota=2
    \footnote{$^{\nttsymb}$}{#1}
    \else\footnote{$^{\the\nmbnota}$}{#1}\fi}\fi}
\def\nttlett{\ifcase\nmbnota \or a\or b\or c\or d\or e\or f\or
g\or h\or i\or j\or k\or l\or m\or n\or o\or p\or q\or r\or
s\or t\or u\or v\or w\or y\or x\or z\fi}
\def\nttsymb{\ifcase\nmbnota \or\dag\or\sharp\or\ddag\or\star\or
\natural\or\flat\or\clubsuit\or\diamondsuit\or\heartsuit
\or\spadesuit\fi}   \clearnmbnota
\def\numberfootnote{\global\tipbnota=0} \numberfootnote
\def\setnote#1{\expandafter\xdef\csname#1\endcsname{
\ifnum\tipbnota=1 {\rm\nttlett} \else {\ifnum\tipbnota=2
{\nttsymb} \else \the\nmbnota\fi}\fi} }
\newcount\nbmfig  \def\clearnbmfig{\global\nbmfig=0}
\gdef\figure{\global\advance\nbmfig by 1
      {\rm fig. \the\nbmfig}}   \clearnbmfig
\def\setfig#1{\expandafter\xdef\csname#1\endcsname{fig. \the\nbmfig}}
 \def\endformula{\eqno\numero $$}
 \def\efr{\endformula}
\newcount\frmcount \def\clearfrmcount{\global\frmcount=0}
\def\numero{\global\advance\frmcount by 1   \ifnum\indappcount=0
  {\ifnum\cpcount <1 {\hbox{\rm (\the\frmcount )}}  \else
  {\hbox{\rm (\the\cpcount .\the\frmcount )}} \fi}  \else
  {\hbox{\rm (\applett .\the\frmcount )}} \fi}
\def\nfr{\nameformula}    \def\numali{\numero}
\def\nameformula#1{\global\advance\frmcount by 1%
{\ifnum\indappcount=0%
{\ifnum\cpcount<1\xdef\spzzttrra{(\the\frmcount )}%
\else\xdef\spzzttrra{(\the\cpcount .\the\frmcount )}\fi}%
\else\xdef\spzzttrra{(\applett .\the\frmcount )}\fi}%
\expandafter\xdef\csname#1\endcsname{\spzzttrra}%
\eqno{\ifnum\draftnum=0\hbox{\rm\spzzttrra}\else%
\hbox{$\buildchar{\rm\spzzttrra}{\tt\scriptscriptstyle#1}{}$}\fi}$$}
\def\nameali#1{\global\advance\frmcount by 1%
{\ifnum\indappcount=0%
{\ifnum\cpcount<1\xdef\spzzttrra{(\the\frmcount )}%
\else\xdef\spzzttrra{(\the\cpcount .\the\frmcount )}\fi}%
\else\xdef\spzzttrra{(\applett .\the\frmcount )}\fi}%
\expandafter\xdef\csname#1\endcsname{\spzzttrra}%
\ifnum\draftnum=0\hbox{\rm\spzzttrra}\else%
\hbox{$\buildchar{\rm\spzzttrra}{\tt\scriptscriptstyle#1}{}$}\fi}
\clearfrmcount
\newcount\cpcount \def\clearcpcount{\global\cpcount=0}
\newcount\subcpcount \def\clearsubcpcount{\global\subcpcount=0}
\newcount\appcount \def\clearappcount{\global\appcount=0}
\newcount\indappcount \def\clearindappcount{\indappcount=0}
\newcount\sottoparcount 

\def\applett{\ifcase\appcount  \or {A}\or {B}\or {C}\or
{D}\or {E}\or {F}\or {G}\or {H}\or {I}\or {J}\or {K}\or {L}\or
{M}\or {N}\or {O}\or {P}\or {Q}\or {R}\or {S}\or {T}\or {U}\or
{V}\or {W}\or {X}\or {Y}\or {Z}\fi    \ifnum\appcount<0
\immediate\write16 {Panda ERROR - Appendix: counter "appcount"
out of range}\fi  \ifnum\appcount>26  \immediate\write16 {Panda
ERROR - Appendix: counter "appcount" out of range}\fi}
\clearappcount  \clearindappcount \newcount\connttrre
\def\clearconnttrre{\global\connttrre=0} \newcount\countref
\def\clearcountref{\global\countref=0} \clearcountref
\def\chapter#1{\global\advance\cpcount by 1 \clearfrmcount
                 \goodbreak\null\vbox{\jump\nobreak
                 \clearsubcpcount\clearindappcount
                 \itemitem{\ttaarr\the\cpcount .\qquad}{\ttaarr #1}
                 \par\nobreak\jump\sjump}\nobreak}
\def\section#1{\global\advance\subcpcount by 1 \goodbreak\null
               \vbox{\sjump\nobreak\ifnum\indappcount=0
                 {\ifnum\cpcount=0 {\itemitem{\ppaarr
               .\the\subcpcount\quad\enskip\ }{\ppaarr #1}\par} \else
                 {\itemitem{\ppaarr\the\cpcount .\the\subcpcount\quad
                  \enskip\ }{\ppaarr #1} \par}  \fi}
                \else{\itemitem{\ppaarr\applett .\the\subcpcount\quad
                 \enskip\ }{\ppaarr #1}\par}\fi\nobreak\jump}\nobreak}
\clearsubcpcount
\def\appendix#1{\global\advance\appcount by 1 \clearfrmcount
                  \goodbreak\null\vbox{\jump\nobreak
                  \global\advance\indappcount by 1 \clearsubcpcount
          \itemitem{ }{\hskip-40pt\ttaarr Appendix\ \applett :\ #1}
             \nobreak\jump\sjump}\nobreak}
\clearappcount \clearindappcount
\def\references{\goodbreak\null\vbox{\jump\nobreak
   \itemitem{}{\ttaarr References} \nobreak\jump\sjump}\nobreak}

\clearcpcount\clearcountref

\def\setchap#1{\ifnum\indappcount=0{\ifnum\subcpcount=0%
\xdef\spzzttrra{\the\cpcount}%
\else\xdef\spzzttrra{\the\cpcount .\the\subcpcount}\fi}
\else{\ifnum\subcpcount=0 \xdef\spzzttrra{\applett}%
\else\xdef\spzzttrra{\applett .\the\subcpcount}\fi}\fi
\expandafter\xdef\csname#1\endcsname{\spzzttrra}}
\newcount\draftnum \newcount\ppora   \newcount\ppminuti
\global\ppora=\time   \global\ppminuti=\time
\global\divide\ppora by 60  \draftnum=\ppora
\multiply\draftnum by 60    \global\advance\ppminuti by -\draftnum
\def\droggi{\number\day /\number\month /\number\year\ \the\ppora
:\the\ppminuti}     \global\draftnum=0
\def\draftcomment#1{\ifnum\draftnum=0 \relax \else
{\ {\bf ***}\ #1\ {\bf ***}\ }\fi} 
%
%
\catcode`@=11
\gdef\Ref#1{\expandafter\ifx\csname @rrxx@#1\endcsname\relax%
{\global\advance\countref by 1    \ifnum\countref>200
\immediate\write16 {Panda ERROR - Ref: maximum number of references
exceeded}  \expandafter\xdef\csname @rrxx@#1\endcsname{0}\else
\expandafter\xdef\csname @rrxx@#1\endcsname{\the\countref}\fi}\fi
\ifnum\draftnum=0 \csname @rrxx@#1\endcsname \else#1\fi}
\gdef\beginref{\ifnum\draftnum=0  \gdef\Rref{\fairef}
\gdef\endref{\scriviref} \else\relax\fi
\ifx\risposta\mplarisposta \ninepoint \fi
\baselineskip=12pt \parskip 2pt plus.2pt }
\def\Reflab#1{[#1]} \gdef\Rref#1#2{\item{\Reflab{#1}}{#2}}
\gdef\endref{\relax}  \newcount\conttemp
\gdef\fairef#1#2{\expandafter\ifx\csname @rrxx@#1\endcsname\relax
{\global\conttemp=0 \immediate\write16 {Panda ERROR - Ref: reference
[#1] undefined}} \else
{\global\conttemp=\csname @rrxx@#1\endcsname } \fi
\global\advance\conttemp by 50  \global\setbox\conttemp=\hbox{#2} }
\gdef\scriviref{\clearconnttrre\conttemp=50
\loop\ifnum\connttrre<\countref \advance\conttemp by 1
\advance\connttrre by 1
\item{\Reflab{\the\connttrre}}{\unhcopy\conttemp} \repeat}
\clearcountref \clearconnttrre
\catcode`@=12
\ifx\risposta\mplarisposta \def\Reflab#1{#1.} \letterfootnote \fi
%
%

\def\slashchar#1{\setbox0=\hbox{$#1$} \dimen0=\wd0
     \setbox1=\hbox{/} \dimen1=\wd1 \ifdim\dimen0>\dimen1
      \rlap{\hbox to \dimen0{\hfil/\hfil}} #1 \else
      \rlap{\hbox to \dimen1{\hfil$#1$\hfil}} / \fi}
\ifx\oldchi\undefined \let\oldchi=\chi
  \def\cchi{{\raise 1pt\hbox{$\oldchi$}}} \let\chi=\cchi \fi
\ifnum\contasym=1 \else \fi 
 \def\del{\partial}   

\def\frac#1#2{{\textstyle{#1 \over #2}}}

\def\half{\ifinner {\scriptstyle {1 \over 2}}\else {1 \over 2} \fi}

\def\vev#1{\langle#1\rangle}

\def\simge{\rlap{\raise 2pt \hbox{$>$}}{\lower 2pt \hbox{$\sim$}}}
\def\simle{\rlap{\raise 2pt \hbox{$<$}}{\lower 2pt \hbox{$\sim$}}}

\def\buildchar#1#2#3{{\null\!\mathop{#1}\limits^{#2}_{#3}\!\null}}

\def\vbig#1#2{{\vbigd@men=#2\divide\vbigd@men by 2%
\hbox{$\left#1\vbox to \vbigd@men{}\right.\n@space$}}}

\def\noblackbox{\overfullrule=0pt} 
%
%
\newcount\fdefcontre \newcount\fdefcount \newcount\indcount
\newread\filefdef  \newread\fileftmp  \newwrite\filefdef
\newwrite\fileftmp     \def\strip #1*.A {#1}%
\def\futuredef#1{\beginfdef
\expandafter\ifx\csname#1\endcsname\relax%
{\immediate\write\fileftmp{#1*.A}%
\immediate\write16 {Panda Warning - fdef: macro "#1" on page
\the\pageno \space undefined}
\ifnum\draftnum=0 \expandafter\xdef\csname#1\endcsname{(?)}
\else \expandafter\xdef\csname#1\endcsname{(#1)}\fi
\global\advance\fdefcount by 1}\fi\csname#1\endcsname}

\def\beginfdef{\ifnum\fdefcontre=0
\immediate\openin\filefdef\jobname.fdef
\immediate\openout\fileftmp\jobname.ftmp
\global\fdefcontre=1  \ifeof\filefdef \immediate\write16 {Panda
WARNING - fdef: file \jobname.fdef not found, run TeX again}
\else \immediate\read\filefdef to\spzzttrra
\global\advance\fdefcount by \spzzttrra
\indcount=0 \loop\ifnum\indcount<\fdefcount
\advance\indcount by 1%
\immediate\read\filefdef to\spezttrra%
\immediate\read\filefdef to\sppzttrra%
\edef\spzzttrra{\expandafter\strip\spezttrra}%
\immediate\write\fileftmp {\spzzttrra *.A}
\expandafter\xdef\csname\spzzttrra\endcsname{\sppzttrra}%
\repeat \fi \immediate\closein\filefdef \fi}
\def\endfdef{\immediate\closeout\fileftmp   \ifnum\fdefcount>0
\immediate\openin\fileftmp \jobname.ftmp
\immediate\openout\filefdef \jobname.fdef
\immediate\write\filefdef {\the\fdefcount}   \indcount=0
\loop\ifnum\indcount<\fdefcount    \advance\indcount by 1
\immediate\read\fileftmp to\spezttrra
\edef\spzzttrra{\expandafter\strip\spezttrra}
\immediate\write\filefdef{\spzzttrra *.A}
\edef\spezttrra{\string{\csname\spzzttrra\endcsname\string}}
\iwritel\filefdef{\spezttrra}
\repeat  \immediate\closein\fileftmp \immediate\closeout\filefdef
\immediate\write16 {Panda Warning - fdef: Label(s) may have changed,
re-run TeX to get them right}\fi}
\def\iwritel#1#2{\newlinechar=-1
{\newlinechar=`\ \immediate\write#1{#2}}\newlinechar=-1}
\global\fdefcontre=0 \global\fdefcount=0 \global\indcount=0
%
%
%
\mathchardef\alpha="710B   \mathchardef\beta="710C
\mathchardef\gamma="710D   \mathchardef\delta="710E
\mathchardef\epsilon="710F   \mathchardef\zeta="7110
\mathchardef\eta="7111   \mathchardef\theta="7112
\mathchardef\iota="7113   \mathchardef\kappa="7114
\mathchardef\lambda="7115   \mathchardef\mu="7116
\mathchardef\nu="7117   \mathchardef\xi="7118
\mathchardef\pi="7119   \mathchardef\rho="711A
\mathchardef\sigma="711B   \mathchardef\tau="711C
\mathchardef\upsilon="711D   \mathchardef\phi="711E
\mathchardef\chi="711F   \mathchardef\psi="7120
\mathchardef\omega="7121   \mathchardef\varepsilon="7122
\mathchardef\vartheta="7123   \mathchardef\varpi="7124
\mathchardef\varrho="7125   \mathchardef\varsigma="7126
\mathchardef\varphi="7127
%
%
\null
%
%
%
%
%
\loadamsmath
\noblackbox
\def\Teta#1#2{\Theta\left[{}^{#1}_{#2}\right]}
\def\di{{\rm d}}
\nopagenumbers
{\baselineskip=12pt
\line{\hfill IFUM-554/FT}
\line{\hfill hep-th/9703066}
\line{\hfill March, 1997}}
{\baselineskip=14pt
\vfill
\centerline{\capsone On the Scattering of Gravitons}
\sjump
\centerline{\capsone on Two Parallel D-Branes}
\bjump\bjump
\centerline{\scaps Andrea Pasquinucci}
\sjump
\centerline{\sl Dipartimento di Fisica, Universit\`a di Milano}
\centerline{\sl and INFN, sezione di Milano}
\centerline{\sl via Celoria 16, I-20133 Milano, Italy}
\bjump \vfill
%
%
\centerline{\capsone ABSTRACT}
\sjump
\noindent
I discuss the scattering of a graviton (or a dilaton,
or an anti-symmetric tensor) on two parallel static Dp-branes.
The graviton belongs to a type II string in 10D.

\sjump \vfill
\pageno=0 \eject }
\yespagenumbers\pageno=1
%
\null\jump
\noindent{\bf 1.~~~Introduction}
\sjump
Many interesting results have been recently obtained from the
computation  of string scattering amplitudes in presence of Dp-branes, 
for a review see for example ref.\ [\Ref{II}] and references therein.
Most of the computations so far appeared in the literature
concern the case of the scattering of a particle on a single 
Dp-brane. The external particle
belongs to a closed string theory, in 10D a type II string,
which interacts with the open Dirichelet string describing the
D-brane.
Thus from a perturbative string
point of view the scattering of, for example, a graviton on a Dp-brane
is described by the amplitude where the two vertex operators of
the incoming and outgoing graviton are inserted in the bulk of a disk.

The next case in perturbation theory is the one in which there are
two parallel Dp-branes and an open string stretching
between them. At first one assumes that the Dp-branes
are fixed in space, or in other words, one completely disregards their
dynamics, even their recoil. The amplitude describing
this scattering is given by the insertion of the two vertex operators
in the bulk of a cylinder which is the world-sheet spanned by the open
string.
More generally, a scattering of a graviton on $n$
parallel Dp-branes is described by a ``$n-1$ loop'' amplitude in
open-string theory with the two graviton vertex operators inserted
in the bulk.

In the case of two parallel Dp-branes, one has to compute
amplitudes with closed string vertex operators inserted
on the cylinder (see for example ref.\ [\Ref{KosteB}]).
The only novelties in the computation are the modifications due to
the fact that the boundary conditions of the open string are Neumann in
$p+1$ directions, and Dirichelet in the remaining directions.

In this letter I will describe this computation
in the case of the scattering of a graviton (or a dilaton or an
anti-symmetric tensor) belonging to a D=10 type II string theory,
on two parallel Dp-branes.

Before getting to the computation, it is worth to discuss some issues.
As already said, the Dp-branes are assumed to be fixed in space and 
indeed in the computation I will completely ignore their dynamics, 
even their recoil. 
Whereas on the disc, i.e.\ for the scattering on a
single Dp-brane, this did not lead to any pathology, on the cylinder
a priori, the non-conservation of the momentum in the direction orthogonal 
to the branes, could lead to some ``inconsistencies'' in the final result.

Indeed, in the explicit computation, the external particles are on-shell
but the momentum is not conserved since part of it is absorbed by the
Dp-branes. This situation can be thought of as if some particles
were off-shell, and indeed Dirichelet boundary condition were introduced
some time ago as a way of computing off-shell string scattering amplitudes
[\Ref{Green},\Ref{Cohen}]. But it is very well-known how subtle is
the process of ``going off-shell'' in a string scattering amplitude since
it also means violating the 2d conformal invariance on which the
perturbative formulation of the scattering amplitude is based.

For example, in the presence of the Dp-brane and since the momentum is 
not fully conserved, one could be worried that the final
expression of the amplitude could cease to be independent from the
insertion points of the Picture Changing Operators, could be not gauge
invariant and there could exist some ``conformal anomalies'' 
[\Ref{Cohen}].
This implies that the computation could be plagued by ambiguities and
the result must then be handled with care. 

Of course, the best approach would be to take in consideration also 
the dynamics of the D-branes. But since this is not yet 
possible, one can reverse the argument and from a detailed study of 
the properties of this and similar amplitudes, try to learn something
on the perturbative dynamics (at least of the recoil) of the D-branes 
[\Ref{Billo},\Ref{Tobe}].

\sjump
\noindent{\bf 2.~~~Generalities on the Scattering}
\sjump
Consider two parallel Dp-branes at distance $\Delta Y_\mu$ and an open
string which connects them. As usual the open string fields $X_\mu$,
$\psi_\mu$ have Neumann boundary conditions for $\mu=0,\ldots,p$ and
Dirichelet boundary conditions for $\mu=p+1,\ldots,9$. The scattering of
a graviton (or a Kalb-Ramond tensor or a dilaton) of a 10D type II
string on the two parallel Dp-branes, i.e.\ on the open string which
connects them, is then described by an amplitude on a cylinder with the
two closed string vertex operators, describing the incoming and outgoing
particles, inserted in the bulk.

I work in the usual string formalism following mostly ref.\ [\Ref{KosteB}],
my conventions for the prime form and the theta functions are as in
refs.\ [\Ref{ammedm},\Ref{mink},\Ref{normaliz}]. For what concerns the
Dp-brane, I follow mostly the conventions and notations of ref.\
[\Ref{Garousi}]. I use the NSR formalism and I am fully covariant (all
ghosts and superghosts). This formalism is the most practical for
extending these computations also to scattering of space-time fermions
and other particles.

To fix the notation I first recall the form of the partition function.

\sjump
\noindent{2.1~~{\it The Partition function}}
\sjump
The partition function (see [\Ref{Polc}]) for a Dp-brane is
$$
\eqalignno{
Z_p\ =\ & N^p_1 \int_{0}^{\infty} \di \Im{}m\tau\ \times\
{(k)^{-26/24} (\eta(\tau))^2 \over \Im{}m\tau}\ \times\ &\numali\cr
&e^{-(\Delta Y)^2 \Im{}m\tau /4\pi} (k)^{10/24} (\eta(\tau))^{-10}
(2\pi \Im{}m\tau)^{-(p+1)/2}\ \times\ \cr
& \sum_{\alpha,\beta} C_\beta^\alpha
\left( \Teta\alpha\beta(0\vert\tau)\right)^5
(k)^{5/24} (\eta(\tau))^{-5} \ \times\
\left( \Teta\alpha\beta(0\vert\tau) \right)^{-1}
(k)^{1/2-1/24} \eta(\tau) \cr}
$$
where I divided the contribution of $bc$-ghosts, bosonic coordinates,
fermionic coordinates and $\beta\gamma$-superghosts respectively.
In this formula $k=\exp(2\pi i\tau)$,
$N_1^p = (4\pi\alpha')^{-(p+1)/2}$ is the overall
normalization, the sum is over only the even spin-structures (the odd
spin-structure does not contribute) of the open string
where $\alpha=0$ is Ramond (R) and $\alpha=1/2$ is Neveu-Schwartz (NS),
and $C^\alpha_\beta= \frac12 \exp[2\pi i(\alpha+\beta)]$. Moreover, on
the cylinder $\Re{}e\tau=0$.

The partition function vanishes by summing over the spin-structures.

\sjump
\noindent{\bf 3.~~~The two-point amplitude}
\sjump
The vertex operator for a graviton (or a anti-symmetric tensor or a
dilaton) in the zero super-ghost picture is
$$
\eqalignno{
V^{Dp}(\bar{z},z;k,\zeta) \ =\ \zeta^{\rho_1\nu_2}{\kappa\over \pi} \,
& [\bar\del \bar{X}_{\rho_1} (z,\bar{z}) -i k\cdot \bar\psi(\bar{z})
\bar\psi_{\rho_1}(\bar{z})] \ \times\cr
& [\del X_{\nu_2} (z,\bar{z}) -i k\cdot\psi(z)\psi_{\nu_2}(z)]
e^{ik\cdot X(z,\bar{z})}
&\numali\cr}
$$
where $k^\mu=\sqrt{\alpha'/2}\, p^\mu$, $k^2=0$, 
and the polarization tensor satisfies $k\cdot\zeta=\zeta\cdot k=0$.
The amplitude is given by
$$
\eqalignno{
T^{(Dp)}(k_1,\zeta_1;k_2,\zeta_2)\ =\ & \int_{0}^{\infty}
{\di \Im{}m\tau \over \Im{}m\tau}\,  \int \di^2 z_1 \di^2 z_2
&\nameali{ampl}\cr
& (8\pi^2 \alpha' \Im{}m\tau)^{-(p+1)/2}
(\eta(\tau))^{-12} e^{-(\Delta Y)^2 \Im{}m\tau /4\pi} \cr
&\sum_{\alpha,\beta} C_\beta^\alpha
\left( \Teta\alpha\beta(0\vert\tau)\right)^4
\vev{V^{Dp}(\bar{z}_1,z_1;k_1,\zeta_1)
V^{Dp}(\bar{z}_2,z_2;k_2,\zeta_2) }\ .\cr}
$$
In my conventions, I
extract from all correlators the contribution of the partition function so
that the amplitude is given by the partition function times the correlator
of the vertex operators. I am using vertex operators in the
zero-superghost picture, so that the contribution to the amplitude
of the superghosts is exactly the same as in the partition function
and there is no need for the insertion of Picture Changing Operators.

I am not making any assumptions on the polarizations so that the
result holds for gravitons, Kalb-Ramond tensors and dilatons.

To proceed in the computation of eq.\ \ampl\ one notices immediately
that the odd spin-structure does not contribute, since to have a
contribution from the odd spin-structure one needs to have at least
ten world-sheet fermions.~\note{This of course holds in 10 dimensions
and for vertex operators in the zero super-ghost picture. In other
dimensions or using different pictures, similar, but not identical,
results hold.}
Then the sum is restricted to the even spin-structures.
By the Riemann (or abstruse) identity, only terms
with eight, or more, world-sheet fermions give a non vanishing
contribution after the sum over the spin-structures.
Thus {\it effectively\/} I can use the following vertex
operator~\note{I stress the fact that this result holds only for this
computation and in the zero super-ghost picture.}
$$
V^{Dp,eff}(\bar{z},z;k,\zeta) \ =\ -\zeta^{\rho_1\nu_2}{\kappa\over \pi} \,
k\cdot \bar\psi(\bar{z})\bar\psi_{\rho_1}(\bar{z})
\ k\cdot\psi(z)\psi_{\nu_2}(z) e^{ik\cdot X(z,\bar{z})}\ .
\efr
As in ref.\ [\Ref{Garousi}] I introduce the projectors $V$ parallel to the
Dp-brane and $N$ orthogonal to the Dp-brane. Thus
$g_{\mu\nu}=V_{\mu\nu}+N_{\mu\nu}$ and $D_{\mu\nu}=V_{\mu\nu}-N_{\mu\nu}$
with $g=(-1,+1,\ldots,+1)$. Notice also that
$D_{\mu}^{~\lambda} D_{\lambda\nu}=g_{\mu\nu}$.
Momentum is conserved only in the directions parallel to the Dp-branes,
i.e.\ $\sum_i (V\cdot k_i)=0$, which for a two-point amplitude
is equivalent to $k_1 + D\cdot k_1 + k_2 + D\cdot k_2 =0$.
Finally $t=-(k_1+k_2)^2=-2k_1\cdot k_2$ is the momentum transferred to the
Dp-brane, and $q^2=k_1\cdot V\cdot k_1=\frac12 k_1\cdot D\cdot k_1$ is the
momentum flowing parallel to the world-volume of the brane.

With these notations, the effective vertex operator can be written as
$$
V^{Dp,eff}(\bar{z},z;k,\zeta) \ =
-{\kappa\over \pi} (D\cdot\zeta)^{\nu_1\nu_2} (k\cdot D)^{\mu_1}
k^{\mu_2}\psi_{\mu_1}(\bar{z})\psi_{\nu_1}(\bar{z})
\psi_{\mu_2}(z)\psi_{\nu_2}(z) e^{ik\cdot X(z,\bar{z})}\, .
\nfr{effvert}
So what is left to do, is to compute the bosonic and fermionic correlators
in eq.\ \ampl. 

Similar bosonic correlators have appeared already in the literature, see
for example refs.\
[\Ref{Cohen},\Ref{Giddings},\Ref{Burgess},\Ref{Blau},\Ref{Li},\Ref{Gutperle},\Ref{Pesando},\Ref{Iengo}].
Since 
I am not integrating over the distance
$\Delta Y$ between the two Dp-branes, the result is 
$$\eqalignno{
\vev{e^{ik_1\cdot X(z_1,\bar{z}_1)} & e^{ik_2\cdot X(z_2,\bar{z}_2)} }\ =\
\prod_{i=1}^2 \exp\left[i k_i\cdot N\cdot Y + {1\over 2\pi}k_i\cdot
N\cdot \Delta Y \log\left({z_i\over\bar{z}_i}\right) \right]\ \times \cr
&\exp\left[k_1\cdot k_2
\left(g_X(z_1,z_2) + g_X(\bar{z}_1,\bar{z}_2)\right)\ +
\right.\cr &\qquad
k_1\cdot D\cdot k_2
\left(g_X(z_1,\bar{z}_2) +g_X(\bar{z}_1,z_2) \right)\ +\cr
&\qquad\left.
k_1\cdot D\cdot k_1\, g_X(\bar{z}_1,z_1) +
k_2\cdot D\cdot k_2\, g_X(\bar{z}_2,z_2) \right] \cr
=\ &\prod_{i=1}^2 \exp\left[i k_i\cdot N\cdot Y + {1\over 2\pi}k_i\cdot
N\cdot \Delta Y \log\left({z_i\over\bar{z}_i}\right) \right]\ \times \cr
&\exp\left[-{t\over 2}\left(g_X(z_1,z_2)
-g_X(z_1,\bar{z}_2) -g_X(\bar{z}_1,z_2)
+ g_X(\bar{z}_1,\bar{z}_2)\right)\ +\right. \cr
&\qquad \left. 2q^2 \left(g_X(\bar{z}_1,z_1) + g_X(\bar{z}_2,z_2)
-g_X(z_1,\bar{z}_2) -g_X(\bar{z}_1,z_2)\right)\right]
&\nameali{expvevX}\cr}
$$
where
$$
g_X(z_1,z_2)\ =\ g_{osc}(z_1,z_2) +
g_{zero}(z_1,\bar{z}_1;z_2,\bar{z}_2)
\nfr{gxdef}
and
$$\eqalignno{
& g_{osc}(z_1,z_2) \ =\ \log E(z_1,z_2) &\numali\cr
&g_{zero}(z_1,\bar{z}_1;z_2,\bar{z}_2)\ =\  - {1\over 16\pi \Im{}m\tau}
(\log{}z_1 +\log\bar{z}_1 - \log{}z_2 - \log\bar{z}_2)^2 \cr}
$$
(notice that $g_X$ is a function of $(z_1,\bar{z}_1;z_2,\bar{z}_2)$).

The correlator of the world-sheet fermions does not depend on the
position of the Dp-branes, thus the full dependence on $Y$ and $\Delta Y$
is given by
$$
e^{-(\Delta Y)^2 \Im{}m\tau /4\pi}
\,\prod_{i=1}^2 \exp\left[i k_i\cdot N\cdot Y + {1\over 2\pi}k_i\cdot
N\cdot \Delta Y \log\left({z_i\over\bar{z}_i}\right) \right]
\nfr{Ydep}
It is interesting to see what one obtains by integrating out the dependence
on $Y$. In physical terms this means summing up all possible positions
of the Dp-branes and this should establish the full momentum conservation.

Indeed, following for example ref.\ [\Ref{Gutperle}], one can introduce
a momentum $p$ conjugate to $Y$ and a momentum $\hat{p}$ conjugate to
$\Delta Y$ and make a Fourier transform.
The integration over $Y$ gives a $\delta(p+N\cdot k_1+N\cdot k_2)$,
which together with the momentum conservation in the directions
parallel to the Dp-branes gives $\delta(p+k_1+k_2)$.
Thus $p^2=-t$ is the overall momentum transferred
to the system of the Dp-branes from the scattering particles.
$\hat{p}$ is then the momentum of one of the two branes. As
observed in ref.\ [\Ref{Gutperle}], if one chooses to keep one of
the two branes fixed in space so that $\hat{p}=0$ and all the momentum
is transferred to the other, equation \expvevX\ simplifies considerably.

The fermionic correlator in eq.\ \ampl\ is
$$
(D\cdot\zeta)^{\nu_1\nu_2} (k_1\cdot D)^{\mu_1}
k_1^{\mu_2}(D\cdot\zeta)^{\nu_3\nu_4} (k_2\cdot D)^{\mu_3}
k_2^{\mu_4}
\Psi_{\mu_1\nu_1\mu_2\nu_2\mu_3\nu_3\mu_4\nu_4}
\nfr{fcorr}
where
$$
\Psi_{\mu_1\nu_1\mu_2\nu_2\mu_3\nu_3\mu_4\nu_4} =
\vev{\psi_{\mu_1}(\bar{z}_1)\psi_{\nu_1}(\bar{z}_1)
\psi_{\mu_2}(z_1)\psi_{\nu_2}(z_1)
\psi_{\mu_3}(\bar{z}_2)\psi_{\nu_3}(\bar{z}_2)
\psi_{\mu_4}(z_2)\psi_{\nu_4}(z_2) }\ .
\efr
To compute this correlator I use the usual Wick contraction valid
for even spin-structures
$$
\vev{\psi^{\mu_1}(z_1)\psi^{\mu_2}(z_2)}\ =\
{(1+(-1)^S)\over 2} g^{\mu_1\mu_2} {\Teta\alpha\beta(\nu_{12}\vert\tau)
\over E(z_1,z_2)\Teta\alpha\beta(0\vert\tau) }
\efr
where $S=(1-2\alpha)(1+2\beta)$ and 
$\nu_{12}=\frac1{2\pi i}\log{z_1\over z_2}$.

Adding the partition function contribution, a generic term of the
correlator $\Psi$ looks like
$$\eqalignno{
&\sum_{\alpha\beta}C^\alpha_\beta k^{1/2}
\left({k^{1/24}\over \eta(\tau)}\right)^4 {(1+(-1)^S)\over 2}
(-g^{\mu_1\mu_2}g^{\nu_1\mu_3}g^{\nu_2\mu_4}g^{\nu_3\nu_4}) \ \times\cr
&\qquad\qquad\qquad
{\Teta\alpha\beta(\nu_{\bar{1}1}\vert\tau)\Teta\alpha\beta
(\nu_{\bar{1}\bar{2}}\vert\tau)
\Teta\alpha\beta(\nu_{12}\vert\tau)\Teta\alpha\beta(\nu_{\bar{2}2}\vert\tau)
\over E(\bar{z}_1,z_1)E(\bar{z}_1,\bar{z}_2)E(z_1,z_2)E(\bar{z}_2,z_2)}\ = \cr
& -\frac12 (-g^{\mu_1\mu_2}g^{\nu_1\mu_3}g^{\nu_2\mu_4}g^{\nu_3\nu_4})
k^{1/2} \left({k^{1/24}\over \eta(\tau)}\right)^4
\left(\eta(\tau)\right)^{12} \prod_{i=1}^4 \omega(z_i) &\numali\cr}
$$
where in going to the last line I summed over the spin structures and
$\omega(z)=1/z$. Thus one obtains
$$\eqalignno{
&\sum_{\alpha\beta}C^\alpha_\beta k^{1/2}
\left({k^{1/24}\over \eta(\tau)}\right)^4
\left(\Teta\alpha\beta(0\vert\tau)\right)^4
\Psi_{\mu_1\nu_1\mu_2\nu_2\mu_3\nu_3\mu_4\nu_4}\ =\cr
&-\frac12 k^{1/2} \left({k^{1/24}\over \eta(\tau)}\right)^4
\left(\eta(\tau)\right)^{12} \prod_{i=1}^4 \omega(z_i)\ \times\cr
&\qquad\qquad
\left[-g^{\mu_1\mu_2}g^{\nu_1\mu_3}g^{\nu_2\mu_4}g^{\nu_3\nu_4}
+\hbox{\rm 59 (signed) permutations}\right]\ . &\nameali{fermcorr}\cr}
$$
Putting together all terms I obtain
$$
\eqalignno{
T^{(Dp)}(k_1,\zeta_1;k_2,\zeta_2)\ =\ & -{1\over 2}
\left({\kappa\over\pi}\right)^2  {\cal K}(k_1,k_2;\zeta_1,\zeta_2)
\int_{0}^{\infty}
{\di \Im{}m\tau \over \Im{}m\tau} &\nameali{final}\cr
&(8\pi^2 \alpha' \Im{}m\tau)^{-(p+1)/2}
 e^{-(\Delta Y)^2 \Im{}m\tau /4\pi}
\int {\di^2 z_1 \di^2 z_2 \over \bar{z}_1 z_1 \bar{z}_2 z_2 } \cr
&\prod_{i=1}^2 \exp\left[i k_i\cdot N\cdot Y + {1\over 2\pi}k_i\cdot
N\cdot \Delta Y \log\left({z_i\over\bar{z}_i}\right) \right]\ \times \cr
&\exp\left[-{t\over 2}\left(g_X(z_1,z_2)
-g_X(z_1,\bar{z}_2) -g_X(\bar{z}_1,z_2)
+ g_X(\bar{z}_1,\bar{z}_2)\right)\ +\right. \cr
&\qquad \left. 2q^2 \left(g_X(\bar{z}_1,z_1) + g_X(\bar{z}_2,z_2)
-g_X(z_1,\bar{z}_2) -g_X(\bar{z}_1,z_2)\right)\right]\cr}
$$
where the kinematical factor ${\cal K}(k_1,k_2;\zeta_1,\zeta_2)$ comes
entirely from the fermionic correlator and is given by
$$\eqalignno{
-{1\over 2}{\cal K}(k_1,k_2;\zeta_1,\zeta_2) \ =\ & 
\left(-{t\over 2}\right)(2q^2)\left[{\rm tr}(\zeta_1\cdot \zeta_2^T) 
+ {\rm tr}(D\cdot\zeta_1){\rm tr}(D\cdot\zeta_2) 
-{\rm tr}(\zeta_2\cdot D\cdot \zeta_1\cdot D)\right]\cr
&+(2q^2)^2\,{\rm tr}(\zeta_1\cdot\zeta_2^T) + \left(-{t\over 2}\right)^2
{\rm tr}(D\cdot\zeta_1) {\rm tr}(D\cdot\zeta_2)\cr
&+\left(-{t\over 2}\right) \left[{\rm tr}(D\cdot\zeta_2) \{
(k_2\cdot D\cdot \zeta_1\cdot D\cdot k_1) -
(k_1\cdot D\cdot \zeta_1\cdot k_2)\}\right.\cr
&\qquad\qquad +\frac12 (k_2\cdot D\cdot \zeta_2\cdot\zeta_1^T\cdot D\cdot k_1)
+ \frac12 (k_2\cdot D\cdot \zeta_2^T\cdot\zeta_1\cdot D\cdot k_1) \cr 
&\qquad\qquad \left. 
- (k_1\cdot D\cdot \zeta_1\cdot D\cdot \zeta_2\cdot D\cdot k_2) + 
(1\longleftrightarrow 2)\right]\cr
&+(2q^2)\left[{\rm tr}(D\cdot\zeta_2)(k_2\cdot\zeta_1\cdot k_2) -
(k_1\cdot \zeta_2\cdot D\cdot \zeta_1\cdot k_2) \right. \cr
&\qquad\quad -\frac12 (k_2\cdot \zeta_1\cdot \zeta_2^T\cdot D\cdot k_2) +
\frac12(k_2\cdot \zeta_1\cdot \zeta_2^T\cdot D\cdot k_1)  \cr 
&\qquad\quad -\frac12 (k_2\cdot \zeta_1^T\cdot \zeta_2\cdot D\cdot k_2) +
\frac12(k_2\cdot \zeta_1^T\cdot \zeta_2\cdot D\cdot k_1)  \cr 
&\qquad\quad \left. + (1 \longleftrightarrow 2) \right]\ .
&\nameali{kinfact}\cr}
$$

\sjump
\noindent{\bf 4.~~~Comments}
\sjump
A full study of the properties of this amplitude, like its divergencies or
possible conformal ``anomalies'', is left to a future publication. Here I
will just make a few comments, starting from the properties of the 
kinematical factor eq.\ \kinfact.

The kinematical factor is the same as the ``tree level'' one, that is 
the one that it is obtained when a graviton is scattering on a single 
Dp-brane. In particular, it vanishes for a 9-brane 
because $D=+1$ and $t=q^2=0$. This
is what one expects since in this case there is no scattering at all.

If one sets both polarizations orthogonal to the Dp-brane, the kinematical
factor simplifies and for a scattering of a graviton one has
${\cal K}(k_1,k_2;\zeta_1,\zeta_2) \ =\ 
(2q^2)^2 {\rm tr}(\zeta_1\cdot\zeta_2)$
and for an anti-symmetric tensor
${\cal K}(k_1,k_2;\zeta_1,\zeta_2) \ =\ (2q^2)[
4(k_1\cdot \zeta_2\cdot \zeta_1 k_2)+
(t-2q^2){\rm tr}(\zeta_1\cdot \zeta_2)]$
whereas for an incoming graviton and an outgoing anti-symmetric tensor
the kinematical factor vanishes.~\note{A different model where this 
contribution does not vanish is studied in ref.~[\Ref{Iengo}].}

The way in which the amplitude is formulated easily allows to study 
various properties of this Dp-brane system. For example, one can 
consider the case where the two Dp-branes are fixed in space. One can then
freely choose $Y=0$ and study the behaviour of this amplitude as 
a function of $\Delta Y$. It will be of particular interest to discuss the 
limit $\Delta Y \rightarrow 0$, that is the limit in which the 
two Dp-branes coincide in space, in the case of the scattering
of Ramond-Ramond states on the Dp-brane [\Ref{Tobe}]. 

On the other side, as already mentioned, one can integrate over $Y$ and 
$\Delta Y$ introducing the corresponding conjugate momenta. Setting 
$\hat{p}=0$, one can study the behaviour of this scattering amplitude 
as a function of the external momenta $k_1,k_2,p$. For example, one can
show that the divergencies which appear when the vertex operators approach
the boundary of the cylinder are a result of the expected analytical structure
of the amplitude. This contrasts with the case discussed in ref.\ 
[\Ref{Cohen}], where it was found a divergence which led to a Weyl anomaly, 
and ref.\ [\Ref{Gutperle}].

\sjump
\noindent{\bf Acknowledgements}
\sjump
I am indebted to Igor Klebanov for suggestions, discussions and critical 
comments. I thank L.~Girardello and R.~Iengo for discussions. 

This work is partially supported by the European Commission TMR programme
ERBFMRX-CT96-0045 in which A.P.\ is associated to the Milano University.

\references
\beginref
\Rref{mink}{A.~Pasquinucci and K.~Roland, Phys.\ Lett. {\bf B351} (1995) 131
\hbox{[hep-th/9503040]}.}
\Rref{normaliz}{A.~Pasquinucci and K.~Roland, Nucl.\ Phys. {\bf B457} (1995)
27 \hbox{[hep-th/9508135]}.}
\Rref{ammedm}{A.~Pasquinucci and K.~Roland, Nucl.\ Phys. {\bf B440} (1995) 441
\hbox{[hep-th/9411015]}.}
\Rref{KosteB}{V.A.~Kostelecky, O.~Lechtenfeld and S.~Samuel,
Nucl.\ Phys.\ {\bf B298} (1988) 133.}
\Rref{Polc}{J.~Polchinski, S.~Chaudhuri and C.V.~Johnson, preprint
\hbox{hep-th/9602052},\newline
J.~Polchinski, preprint \hbox{hep-th/9611050}.}
\Rref{Garousi}{M.R.~Garousi and R.C.~Myers, Nucl.\ Phys.\ {\bf B475} (1986)
193 [\hbox{hep-th/9603194}].}
\Rref{Gutperle}{M.~Gutperle, Nucl.\ Phys.\ {\bf B444} (1995) 487
[\hbox{hep-th/9502106}].}
\Rref{Blau}{S.K.~Blau et al., Nucl.\ Phys. {\bf B301} (1988) 285.}
\Rref{Burgess}{C.P.~Burgess and T.R.~Morris, Nucl.\ Phys. {\bf B291} (1987)
256.}
\Rref{Li}{M.~Li, Nucl.Phys. {\bf B420} (1994) 339.}
\Rref{Pesando}{M.~Frau, I.~Pesando, S.~Sciuto, A.~Lerda and R.~Russo,
preprint DFTT~8/97, \hbox{hep-th/9702037}.}
\Rref{Green}{M.B.~Green, Nucl.\ Phys.\ {\bf B124} (1977) 461.}
\Rref{Iengo}{F.~Hussain, R.~Iengo and C.~N\'unez, preprint
IC/97/1, SISSAREF-3/97/EP, \hbox{hep-th/9701143}.}
\Rref{Cohen}{A.~Cohen, G.~Moore, P.~Nelson and J.~Polchinski, Nucl.\ Phys.\
{\bf B281} (1987) 127.}
\Rref{Billo}{M.~Bill\'o, P.~Di~Vecchia and D.~Cangemi, preprint
NBI-HE-97-05, NORDITA~97/7P, \hbox{hep-th/9701190}.}
\Rref{Tobe}{A.~Pasquinucci, to appear.}
\Rref{Giddings}{S.B.~Giddings, preprint UCSBTH-96-29, \hbox{hep-th/9612022}.}
\Rref{II}{A.~Hashimoto and I.~Klebanov, preprint PUPT-1669,
\hbox{hep-th/9611214}.}
\endref
\ciao
%
%